\RequirePackage{amsmath}
\documentclass[runningheads,a4paper]{llncs}
\usepackage[english]{babel}
\usepackage[utf8]{inputenc}
\usepackage{csquotes}
\usepackage{amsmath}
\usepackage{cite}
\usepackage{hyperref}
\usepackage[capitalise,nameinlink]{cleveref}
\crefname{section}{Sect.}{Sect.}
\Crefname{section}{Section}{Sections}
\usepackage{xspace}
\usepackage{centernot}
\usepackage{color}
\usepackage{listings}
\crefname{lstlisting}{listing}{listings}
\Crefname{lstlisting}{Listing}{Listings}
\usepackage{booktabs}
\usepackage[binary,squaren]{SIunits} 
\usepackage{lstautogobble}
\usepackage{lipsum}
\usepackage{xparse}
\usepackage[color=blue!20, linecolor=blue!80!black]{todonotes}
\usepackage{tikz}
\usepackage{dsfont}
\usepackage{listings-rust}
\usetikzlibrary{arrows,automata,positioning,shapes,patterns,arrows.meta}
\usepackage{subcaption}
\usepackage{wrapfig}
\usepackage{pgfplots}
\pgfplotsset{compat=1.8}
\usepgfplotslibrary{statistics}
\usepgfplotslibrary{colorbrewer}
\usetikzlibrary{pgfplots.statistics, pgfplots.colorbrewer}
\usepackage{pgfplotstable}

\newcommand{\donotshow}[1]{}

\definecolor{CommentColor}{rgb}{0.13, 0.13, 1}

\newcommand{\ellplusone}{{\ell + 1}}
\newcommand{\memreqs}{\memreq{s}}
\newcommand\lstmathanno{\color[RGB]{0, 20, 110}}

\definecolor{first}{HTML}{70ACEC}
\definecolor{second}{HTML}{5682B1}
\definecolor{third}{HTML}{3B5979}

\definecolor{time}{HTML}{00A4CC}
\definecolor{mem}{HTML}{F95700}

\definecolor{prelude}{HTML}{F79A3D}
\definecolor{loop}{HTML}{2874ae}
\definecolor{prefix}{HTML}{61B940}
\definecolor{postfix}{HTML}{A370AB}

\DeclareFontFamily{U}{MnSymbolC}{}
\DeclareSymbolFont{MnSyC}{U}{MnSymbolC}{m}{n}
\DeclareFontShape{U}{MnSymbolC}{m}{n}{
    <-6>  MnSymbolC5
   <6-7>  MnSymbolC6
   <7-8>  MnSymbolC7
   <8-9>  MnSymbolC8
   <9-10> MnSymbolC9
  <10-12> MnSymbolC10
  <12->   MnSymbolC12%
}{}
\DeclareMathSymbol{\powerset}{\mathord}{MnSyC}{180}

\newcommand*{\eg}{e.g.\@\xspace}
\newcommand*{\ie}{i.e.\@\xspace}

\newcommand*{\wrt}{with respect to }
\makeatletter
\newcommand*{\etc}{%
  \@ifnextchar{.}%
    {etc}%
    {etc.\@\xspace}%
}
\newcommand*{\cf}{%
  \@ifnextchar{.}%
    {cf}%
    {cf.\@\xspace}%
}
\newcommand*{\etal}{%
  \@ifnextchar{.}%
    {et~al}%
    {et~al.\@\xspace}%
}
\makeatother

\NewDocumentCommand{\twopartdef}{ m m m o }%
{
  \left\{
    \begin{array}{ll}
      #1\quad & \mbox{if } #2 \\
      #3\quad & \IfValueTF{#4}{\mbox{if } #4}{\mbox{otherwise}}
    \end{array}
  \right.
}
\NewDocumentCommand{\threepartdef}{ m m m m m o }%
{
  \left\{
    \begin{array}{ll}
      #1\quad & \mbox{if } #2 \\
      #3\quad & \mbox{if } #4 \\
      #5\quad & \IfValueTF{#6}{\mbox{if } #6}{\mbox{otherwise}}
    \end{array}
  \right.
}

\newcommand*\rtlola{\text{RTLola}\xspace}
\newcommand*\lola{\text{Lola}\xspace}
\newcommand*\rust{\text{Rust}\xspace}
\newcommand*\viper{\text{Viper}\xspace}
\newcommand*\ltl{\textsc{ltl}\xspace}
\newcommand*\fpga{\textsc{fpga}\xspace}

\newcommand*\vhdl{\textsc{vhdl}\xspace}
\newcommand*\copilot{\text{Copilot}\xspace}
\newcommand*\prusti{\text{Prusti}\xspace}
\newcommand*\java{\text{Java}\xspace}

\newcommand*\spec{\varphi}
\newcommand*\preflen[1][\spec]{{\eta^{\leftarrow}_{#1}}} 
\newcommand*\postlen[1][\spec]{{\eta^{\rightarrow}_{#1}}} 
\newcommand*\memcon[1][\spec]{{\memreq^{\ast}_{#1}}}
\DeclareMathOperator{\sizeof}{size}

\makeatletter
\newcommand*\@evalorder{\leq_{\mathit{eo}}}
\newcommand*\evalorder{\mathrel{\@evalorder}}
\newcommand*\extevalorder{\mathrel{\@evalorder^+}}
\makeatother

\newcommand{\val}[2]{v(#1)(#2)}

\newcommand*\naturals{\mathds{N}}

\NewDocumentCommand{\shift}{ s g }{
  \IfNoValueTF{#2}{\Delta}{
    \IfBooleanTF{#1}%
    { \Delta\left({#2}\right) }%
    { \Delta({#2}) }%
  }
}

\NewDocumentCommand{\memreq}{ s g }{
  \IfNoValueTF{#2}{\mu}{
    \IfBooleanTF{#1}%
    { \mu\left({#2}\right) }%
    { \mu({#2}) }%
  }
}
\NewDocumentCommand{\Set}{ s m }{
  \IfBooleanTF{#1}%
  {\{#2\}}%
  {\left\{{#2}\right\}}%
}
\NewDocumentCommand{\card}{ s m }{
  \IfBooleanTF{#1}%
  {\vert#2\vert}%
  {\left\vert{#2}\right\vert}%
}
\newcommand*\given\mid

\newcommand*\layer[1]{\mathit{Layer}(#1)}

\definecolor{bluekeywords}{rgb}{0.13, 0.13, 1}
\definecolor{greentypes}{rgb}{0, 0.5, 0}
\definecolor{redstrings}{RGB}{171, 114, 2}
\definecolor{graynumbers}{rgb}{0.5, 0.5, 0.5}
\definecolor{goldcomments}{rgb}{0.6, 0.4, 0.08}
\definecolor{monitorblue}{RGB}{18, 163, 38}

\lstdefinelanguage{Lola}{
  keywords=[0]{input, output, trigger, constant},
  keywordstyle=[0]\bfseries\color{bluekeywords},
  keywords=[1]{if, then, else, aggregate, defaults, offset, int},
  keywords=[2]{Int8, Int16, Int32, Int64, UInt8, UInt16, UInt32, UInt64, Bool, Float32, Float64, @1Hz, @5Hz, @10Hz, @100mHz, @1kHz},
  keywordstyle=[2]\color{greentypes},
    sensitive=false,
    comment=[l]{//},
    morecomment=[s]{/*}{*/},
    morestring=[b]',
    morestring=[b]"
}

\usetikzlibrary{decorations.pathreplacing,calc}
\newcommand{\tikzmark}[1]{\tikz[overlay,remember picture] \node (#1) {};}
\lstdefinestyle{LolaDefault}{
    language={Lola},
    commentstyle=\color{goldcomments},
    keywordstyle=[1]\color{bluekeywords},
    stringstyle=\color{redstrings},
    numberstyle=\color{graynumbers},
    showstringspaces=false,
    basicstyle=\ttfamily\footnotesize,
}

\begin{document}

\title{Verified Rust Monitors for Lola Specifications\thanks{This work was partially supported by the German Research Foundation~(DFG) as part of the Collaborative Research Center ``Foundations of Perspicuous Software Systems'' (TRR 248, 389792660), and by the European Research Council (ERC) Grant OSARES (No. 683300).}}

\author{%
  Bernd Finkbeiner\orcidID{0000-0002-4280-8441} \and
  Stefan Oswald \and
  Noemi~Passing\orcidID{0000-0001-7781-043X} \and
  Maximilian~Schwenger\orcidID{0000-0002-2091-7575}
}
\institute{%
  CISPA Helmholtz Center for Information Security\\
  66123 Saarbr\"ucken, Germany \\
  \email{\{finkbeiner,noemi.passing,maximilian.schwenger\}@cispa.saarland}\\
  \email{s.oswald@stud.uni-saarland.de}
}
\authorrunning{Finkbeiner et al.}
\maketitle

\begin{abstract}
  The safety of cyber-physical systems rests on the correctness of their monitoring mechanisms.
  This is problematic if the specification of the monitor is implemented
  manually or interpreted by unreliable software.  We present a
  \emph{verifying compiler} that translates specifications given in the stream-based monitoring language \lola to 
  implementations in \rust.  The generated code contains verification
  annotations that enable the \viper toolkit to automatically prove
  functional correctness, absence of memory faults, and guaranteed
  termination.  The compiler parallelizes the evaluation of different
  streams in the monitor based on a dependency analysis of the
  specification.  We present encouraging experimental results obtained with
  monitor specifications found in the literature.
  For every specification, our approach was able to either produce a correctness proof or to uncover errors in the specification.

\end{abstract}

\section{Introduction}\label{sec:intro}

Cyber-physical systems are inherently safety-critical, because 
failures immediately impact the physical environment. A crucial
aspect of the development of such systems is therefore the integration
of reliable monitoring mechanisms.  A \emph{monitor} is a special system
component that typically has broad access to the sensor readings and
the resulting control decisions. The monitor assesses the system's
health by checking its behavior against a specification.  If a
violation is detected, the monitor raises an alarm and initiates
mitigation protocols such as an emergency landing or a graceful
shutdown.

An obvious concern with this approach is that the safety of the system
rests on the correctness of the monitor. \emph{Quis custodiet ipsos custodes?} For simple specifications,
this is not a serious problem. An \ltl~\cite{ltl} specification, for
example, can be translated into a finite-state automaton that is proven to correspond to the semantics of the specification. Implementing such an
automaton correctly as a computer program is not difficult.  For more
expressive specification languages, establishing the
correctness of the monitor is much more challenging. Especially problematic is the use of interpreters, which read the specification as input and then rely on complicated and error-prone software to interpret the specification dynamically at runtime~\cite{rtlolaarxiv,rtlolacavindustrial,rtlolacavtoolpaper,striver,tessla}. Recently,
however, much effort has gone into the development of compilers.
Compared to a full-scale interpreter, the code produced by a compiler
for a specific specification is fairly simple and well-structured.
Some compilers even include special mechanisms that increase the
confidence in the monitor.  For example, the \rtlola compiler~\cite{fpgalola} generates \vhdl code that is annotated with tracing information that
relates each line of code back to the specific part of the specification it
implements.  The \copilot compiler \cite{copilot} produces a test suite for the generated C code. The framework even includes a bounded
model checker, which can check the correctness of the output
for input sequences up to a fixed length.  However, none of these approaches
actually proves the functional correctness of the monitor.

In this paper, we present a
  \emph{verifying compiler} that translates specifications given in the stream-based monitoring language \lola~\cite{lola} to 
  implementations in \rust\footnote{\url{https://www.rust-lang.org/}}. The generated
code is fully annotated with formal function contracts, loop invariants, and inline assertions, so that functional correctness 
and guaranteed termination can be automatically verified by the
\viper~\cite{viper} toolkit, without any restriction on the length of the
input trace.  Since the memory requirements of a \lola specification
can be computed statically, this yields a formal guarantee that
on any platform that satisfies these requirements, the monitor will
never crash and will always compute the correct output.

A major practical concern for any compiler is the performance of the generated code.
Our \lola-to-\rust compiler produces highly efficient monitor implementations because it parallelizes the code for the evaluation of the specifications.
Since \lola is a stream-based specification language, it exhibits a highly modular and memory-local structure, \ie, the computation of a stream writes only in its own local memory, although it may read from the local memory of several other processes.
The compiler statically analyzes the dependencies between the streams, resulting in a partial evaluation order. To prove correctness, it is shown that streams that are incomparable with respect to the evaluation order can indeed be evaluated in parallel.

We have used our compiler to build monitors from specifications of varying sizes found in the literature. In our experience, the compiler itself scales very well.
The verification in \viper, however, is expensive. It appears that the running times of the underlying 
\textsc{smt} solver Z3~\cite{z3} vary greatly, even for different runs on the same monitor and specification.
Nevertheless, we have been successful in all our benchmarks in the sense that the compiler either generated a verified monitor or uncovered an error in the specification.
This is a major step forward towards the \emph{verified monitoring} of real-life safety-critical systems.

\section{Introduction to Lola}\label{sec:compilation}

The source language of our verifying compiler is the stream-based monitoring language \lola~\cite{lola}.
A \lola monitor is a reactive component that translates, in an online fashion, input streams into output streams. In each time step, the monitor receives new values for the input streams and produces new values for the output streams in accordance with the specification. In principle, the monitoring can continue forever; if the monitor is terminated, it wraps up the remaining computations, produces a final output, and shuts down. \lola specifications are declarative in the sense that the semantics leaves a lot of implementation freedom: the semantics defines how specific values are combined arithmetically and logically, but the precise evaluation order and the memory management are determined by the implementation. 

A \lola specification defines a set of streams.
Each stream is an ordered sequence of typed values that is extended throughout the monitor execution.
There are three kinds of streams:
\begin{description}
  \item[Input Streams] constitute the interface between the monitor and an external data source, \ie, the system under scrutiny.
  \item[Output Streams] compute new values based on input streams, other output streams, and constant values.  The computed values contain relevant information regarding the performance and health status of the system.
  \item[Triggers] constitute the interface between the monitor and the user.  Trigger values are binary and indicate the violation of a property.  In this case, the monitor alerts the user.
\end{description}

Syntactically, a \lola specification is given as a sequence of stream declarations.
Input stream declarations are of the form $i_j: T_j$, where $i_j$ is an input stream and $T_j$ is its type.
Output stream and trigger declarations are of the form $s_j : T_j = e_j(i_1,\dots,i_m,s_1,\dots,s_n)$, where $i_1, \dots, i_m$ are input streams, $s_1, \dots, s_n$ are output streams, and the $e_j$ are stream expressions.
A stream expression consists of constant values, streams, arithmetic and logic operators $f(e_1, \dots, e_k)$, if-then-else expressions \texttt{ite}$(b,e_1,e_2)$, and stream accesses $e[k,c]$, where $e$ is a stream, $k$ is the \emph{offset}, and $c$ is the constant \emph{default value}.
Stream accesses are either \emph{synchronous}, \ie, a stream accesses the latest value of a stream, or \emph{asynchronous}, \ie, a stream accesses a past or future value of another stream.

\begin{figure}[t]
\begin{lstlisting}[style=LolaDefault,caption={A \lola specification monitoring the altitude of a drone. The output stream \texttt{tooLow} (\texttt{tooHigh}) checks whether the drone flies below (above) a given minimum (maximum) altitude in the last, current, and next step. If this is the case, an alarm is raised.}, label=fig:runningexample:spec]
input altitude: Int32
output tooLow: Bool := 
  altitude[-1,0] < 200 & altitude < 200 & altitude[1,0] < 200
output tooHigh: Bool := 
  altitude[-1,0] > 600 & altitude > 600 & altitude[1,0] > 600
trigger tooLow "Flying below minimum altitude."
trigger tooHigh "Flying above maximum altitude."
\end{lstlisting}
\end{figure}

The example specification shown in \Cref{fig:runningexample:spec}  monitors the altitude of a drone, detects whether the drone flies below a given minimum altitude or above a given maximum altitude for too long, and raises an alarm if needed. 
The input stream \texttt{altitude} contains sensor information of the drone. 
The output stream \texttt{tooLow} checks whether the altitude is lower than the given minimum altitude of \texttt{200} in the last, current, and next step, denoted by \texttt{altitude[-1,0]}, \texttt{altitude}, and \texttt{altitude[1,0]}, respectively.
If this is the case, a trigger is raised.
Analogously, \texttt{tooHigh} checks whether the altitude is above the given maximum altitude in the last, current, and next step, and a trigger is raised in this case.
The evaluations of \texttt{tooHigh} and \texttt{tooLow} try to access the second to last value of \texttt{altitude} as well as the last and the next one.
If \texttt{altitude} does not have at least two values, the accesses with offset $-1$ fail and the default value, in this case $0$, is used. 
If \texttt{altitude} ceases to produce values, the accesses with offset $1$ fail. Hence, in contrast to negative offsets, the default value for accesses with positive offset is used at the end of the execution.

The semantics of \lola is defined in terms of \emph{evaluation models}. 
Intuitively, an evaluation model consists of evaluations of each output stream of the specification. The evaluation is a natural translation of the stream expressions. The full formal definition is given in~\cite{lola}.

\begin{definition}[Evaluation Model~\cite{lola}]
	Let $\varphi$ be a \lola specification over input streams $i_1, \dots, i_\ell$ and output streams $s_1, \dots, s_n$. The tuple $\langle\sigma_1, \dots, \sigma_n\rangle$ of streams of length $N+1$ is called an \emph{evaluation model} if for each equation $s_j = e_j(i_1, \dots,i_\ell,s_1,\dots,s_n)$ in $\varphi$, $\langle\sigma_1, \dots, \sigma_n\rangle$ satisfies $\sigma_j(k) = \val{e_j}{k}$ for $0 \leq k \leq N$, where $\val{e_j}{k}$ evaluates the stream expression $e_j$ at position $k$.
\end{definition}

\begin{figure}[t]
	\begin{subfigure}[b]{0.48\textwidth}
		\begin{center}
		\scalebox{0.91}{
		\begin{tikzpicture}[->,shorten >=0pt,auto,stream/.style={draw,minimum height=15pt, minimum width=18pt},time/.style={draw=none, minimum width=18pt}]
		
				\node[]				(t)			at (-0.7,0.8)				{\scriptsize$t$};
				\node[]				(b)			at (-0.7,0)					{$b$};
				\node[]				(a)			at (-0.7,-1.2)				{$a$};
		
				\node[time]			(t-1)		at (0,0.8)					{\scriptsize$-1$};
				\node[time]			(t0)			[right=-0.5 pt of t-1]		{\scriptsize$0$};
				\node[time]			(t11)		[right=-0.5 pt of t0]		{\scriptsize$1.1$};
				\node[time]			(t12)		[right=-0.5 pt of t11]		{\scriptsize$1.2$};
				\node[time]			(t21)		[right=-0.5 pt of t12]		{\scriptsize$2.1$};
				\node[time]			(t22)		[right=-0.5 pt of t21]		{\scriptsize$2.2$};
				\node[time]			(t31)		[right=-0.5 pt of t22]		{\scriptsize$3.1$};
				\node[time]			(t32)		[right=-0.5 pt of t31]		{\scriptsize$3.2$};
				\node[]				(td)			[right=-0.5 pt of t32]		{\scriptsize$\dots$};
	
				\node[stream]		(b-1)		at (0,0)						{$-$};
				\node[stream]		(b0)			[right=-0.5 pt of b-1]		{$-$};
				\node[stream]		(b11)		[right=-0.5 pt of b0]		{$1$};
				\node[stream]		(b12)		[right=-0.5 pt of b11]		{};
				\node[stream]		(b21)		[right=-0.5 pt of b12]		{$2$};
				\node[stream]		(b22)		[right=-0.5 pt of b21]		{};
				\node[stream]		(b31)		[right=-0.5 pt of b22]		{$3$};
				\node[stream]		(b32)		[right=-0.5 pt of b31]		{};
				\node[]				(bd)			[right=-0.5 pt of b32]		{$\dots$};
				
				\node[stream]		(a-1)		at (0,-1.2)					{$-$};
				\node[stream]		(a0)			[right=-0.5 pt of a-1]		{$-$};
				\node[stream]		(a11)		[right=-0.5 pt of a0]		{};
				\node[stream]		(a12)		[right=-0.5 pt of a11]		{$1$};
				\node[stream]		(a21)		[right=-0.5 pt of a12]		{};
				\node[stream]		(a22)		[right=-0.5 pt of a21]		{$2$};
				\node[stream]		(a31)		[right=-0.5 pt of a22]		{};
				\node[stream]		(a32)		[right=-0.5 pt of a31]		{$3$};
				\node[]				(ad)			[right=-0.5 pt of a32]		{$\dots$};
				
				\path	(b11)	edge[thick,first,in=70,out=110]		node		{}	(b0)
						(b21)	edge[thick,second,in=60,out=120]		node		{}	(b11)
						(b31)	edge[thick,third,in=60,out=120]		node		{}	(b21)
						(a12)	edge[thick,first,in=270,out=90]		node		{}	(b11)
						(a22)	edge[thick,second,in=270,out=90]		node		{}	(b21)
						(a32)	edge[thick,third,in=270,out=90]		node		{}	(b31);
			\end{tikzpicture}}
		\end{center}
		\vspace{-5pt}
		\caption{The result of evaluating the output streams respecting the evaluation order.}\label{fig:dependencychangesmodel:correct}
	\end{subfigure}
	\hfill
	\begin{subfigure}[b]{0.48\textwidth}
		\begin{center}
		\scalebox{0.91}{
		\begin{tikzpicture}[->,shorten >=0pt,auto,stream/.style={draw,minimum height=15pt, minimum width=18pt},time/.style={draw=none, minimum width=18pt}]
		
				\node[]				(t)			at (-0.7,0.8)				{\scriptsize$t$};
				\node[]				(a)			at (-0.7,0)					{$a$};
				\node[]				(b)			at (-0.7,-1.2)				{$b$};
		
				\node[time]			(t-1)		at (0,0.8)					{\scriptsize$-1$};
				\node[time]			(t0)			[right=-0.5 pt of t-1]		{\scriptsize$0$};
				\node[time]			(t11)		[right=-0.5 pt of t0]		{\scriptsize$1.1$};
				\node[time]			(t12)		[right=-0.5 pt of t11]		{\scriptsize$1.2$};
				\node[time]			(t21)		[right=-0.5 pt of t12]		{\scriptsize$2.1$};
				\node[time]			(t22)		[right=-0.5 pt of t21]		{\scriptsize$2.2$};
				\node[time]			(t31)		[right=-0.5 pt of t22]		{\scriptsize$3.1$};
				\node[time]			(t32)		[right=-0.5 pt of t31]		{\scriptsize$3.2$};
				\node[]				(td)			[right=-0.5 pt of t32]		{\scriptsize$\dots$};
	
				\node[stream]		(a-1)		at (0,0)						{$-$};
				\node[stream]		(a0)			[right=-0.5 pt of a-1]		{$-$};
				\node[stream]		(a11)		[right=-0.5 pt of a0]		{$1$};
				\node[stream]		(a12)		[right=-0.5 pt of a11]		{};
				\node[stream]		(a21)		[right=-0.5 pt of a12]		{$1$};
				\node[stream]		(a22)		[right=-0.5 pt of a21]		{};
				\node[stream]		(a31)		[right=-0.5 pt of a22]		{$2$};
				\node[stream]		(a32)		[right=-0.5 pt of a31]		{};
				\node[]				(ad)			[right=-0.5 pt of a32]		{$\dots$};
				
				\node[stream]		(b-1)		at (0,-1.2)					{$-$};
				\node[stream]		(b0)			[right=-0.5 pt of b-1]		{$-$};
				\node[stream]		(b11)		[right=-0.5 pt of b0]		{};
				\node[stream]		(b12)		[right=-0.5 pt of b11]		{$1$};
				\node[stream]		(b21)		[right=-0.5 pt of b12]		{};
				\node[stream]		(b22)		[right=-0.5 pt of b21]		{$2$};
				\node[stream]		(b31)		[right=-0.5 pt of b22]		{};
				\node[stream]		(b32)		[right=-0.5 pt of b31]		{$3$};
				\node[]				(bd)			[right=-0.5 pt of b32]		{$\dots$};
				
				\path	(b12)	edge[thick,first,in=60,out=120]		node		{}	(b0)
						(b22)	edge[thick,second,in=60,out=120]		node		{}	(b12)
						(b32)	edge[thick,third,in=60,out=120]		node		{}	(b22)
						(a11)	edge[thick,first,in=90,out=270]		node		{}	(b0)
						(a21)	edge[thick,second,in=90,out=270]		node		{}	(b12)
						(a31)	edge[thick,third,in=90,out=270]		node		{}	(b22);
			\end{tikzpicture}}
		\end{center}
		\vspace{-5pt}
		\caption{The result of evaluating the output streams in order of their declaration.}\label{fig:dependencychangesmodel:wrong}
	\end{subfigure}
	\vspace{5pt}
	\caption{Two different evaluations of the output streams $a$ and $b$, where $a$ accesses~$b$ synchronously and $b$ accesses its previous value. Both accesses default to $0$ and both $a$ and $b$ increase the obtained value by $1$.}\label{fig:dependencychangesmodel}
\end{figure}
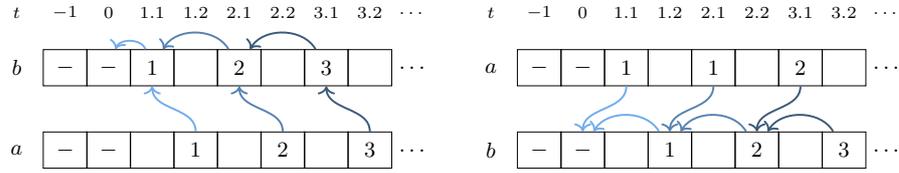

Synchronous accesses harbor a pitfall for the monitor realization as illustrated in \Cref{fig:dependencychangesmodel}. Consider the corresponding \lola specification:
\begin{lstlisting}[style=LolaDefault]
output a: Int32 := b[ 0, 0] + 1
output b: Int32 := b[-1, 0] + 1
\end{lstlisting}  

Here, $a$ accesses $b$ synchronously, while $b$ accesses its previous value.
The evaluation of $a$ tries to access the current value of $b$ and increases the result by one, which yields the next stream value of $a$.
In contrast, the evaluation of $b$ tries to access the last value of $b$ and increases the result by one to determine the next stream value of $b$.
\Cref{fig:dependencychangesmodel:correct} depicts the resulting output.
If the monitor evaluates the streams in order of their declaration, however, the resulting output, shown in \Cref{fig:dependencychangesmodel:wrong}, differs from the expected one.
The reason is that the \emph{current} value of $b$ changes depending on whether or not $b$ has already been extended when accessing the value.
This problem is solved by respecting the evaluation order, a partial order on the output streams.
It is induced by the dependency graph of a \lola specification.

\begin{definition}[Dependency Graph~\cite{lola}]
  The \emph{dependency graph} $D_\spec = (V, E)$ of a \lola specification $\spec$ is a weighted directed multigraph.  Each vertex represents a stream and each edge an access operation.
  Thus, $s \in V$ iff $s$ is a stream or trigger in $\spec$ and $(s_1, n, s_2) \in E$ for $s_1, s_2 \in V$, $n \in \naturals$ iff the stream expression of $s_1$ contains an access to $s_2$ with offset $n$.
\end{definition}

Based on the dependency graph, d'Angelo~\etal define the \emph{shift} of a stream~\cite{lola}. Intuitively, the shift of $s$ indicates how many steps the evaluation of its expression needs to be delayed.
For instance, suppose the delay is $n > 0$. 
Then the value of $s$ for time $t$ can be computed at time $t+n$.

\begin{definition}[Shift~\cite{lola}]
For a \lola specification $\spec$, the \emph{shift} $\shift{s}$ of a stream $s$ is the greatest weight of a path through the dependency graph of $\spec$ originating in $s$:
$
  \shift{s} = \max(0, \max \Set{w + \shift{s'} | (s, w, s') \in E}).
$
\end{definition}

The shift allows us to define an order in which streams need to be evaluated. 
For this, we define the set of synchronized edges $E^\ast$ where the weight of a synchronized edge $(s, n, s') \in E^\ast$ indicates when $s$ can access $s'$ successfully with an offset of $n$.
Let $E^\ast = \Set{(s, \shift{s} - w - \shift{s'}, s') \given (s, w, s') \in E}$.

\begin{definition}[Evaluation Order]
	The \emph{evaluation order} $\evalorder$ is a partial order on the output streams of a \lola specification $\spec$. Let $D_\spec=(V,E)$ be the dependency graph of $\spec$. 
	The evaluation order is the transitive closure of a relation $\prec$ with $s \prec s'$ iff $(s', 0, s) \in E^\ast$.
\end{definition}

Clearly, we obtain $b \evalorder a$ for the above \lola specification, yielding the expected result depicted in \Cref{fig:dependencychangesmodel:correct}.
For the \lola specification from \Cref{fig:runningexample:spec}, however, the output streams \texttt{tooLow} and \texttt{tooHigh} are incomparable according to the evaluation order.
A total evaluation order on the output streams, denoted~$\extevalorder$, is obtained by relating incomparable streams arbitrarily.

\begin{remark}[On Asynchronous Accesses and Off-by-one Errors]
It is fairly easy to make off-by-one errors in asynchronous stream accesses.  
When two streams within one layer access each other asynchronously, one of the offsets needs to be decreased by 1, depending on which stream is evaluated first. This cannot be avoided for any $\extevalorder$. 
To simplify the presentation, we will ignore this issue in the remainder of the presentation, the correct adjustment of the indices is, however, implemented in the compiler.
\end{remark}

Specifications where the dependency graph has no positive cycles are called \emph{efficiently monitorable}: such specifications can be monitored with constant memory, and an output value can always be produced after a constant delay~\cite{lola}. All example specifications considered in this paper are efficiently monitorable.

\section{From Lola to Rust}

The compilation proceeds in two steps.
First, the \lola specification is analyzed to determine inter-stream dependencies, the overall memory requirement, and the different phases of the monitoring process.
Second, the compiler produces \rust code that implements the specification.

\subsection{Specification Analysis}

\paragraph*{Execution Pre- and Postfix.}
Refer back to the \lola specification in \Cref{fig:runningexample:spec}. Another beneficial property of the synchronous input model is that, starting from $t=2$, both stream accesses with offset $-1$ to \texttt{altitude} will always succeed since the offset refers to the last evaluation of \texttt{altitude} which did already happen at $t \geq 1$.
For a more general analysis, suppose an output stream $s$ accesses another stream $s'$ with an offset of $n$. 
If $n$ is non-positive, then accesses may fail until $t = \shift(s) - n - \shift(s')$, \ie, they will not fail from $\shift(s) - n - \shift(s') + 1$ on.
If $n$ is strictly positive, however, the evaluation of $s$ needs to be delayed by $\shift(s) - n$, \ie, until $s'$ received the respective value.
By generally delaying the execution of $s$, all accesses to $s'$ continue to succeed until $s'$ ceases to produce new values.
As soon as this is the case, the monitor needs to evaluate $s$ for $\shift(s) - n$ more times to compensate for the delay.
For instance, the evaluations of \texttt{tooLow} and \texttt{tooHigh} both have to be delayed by one step.

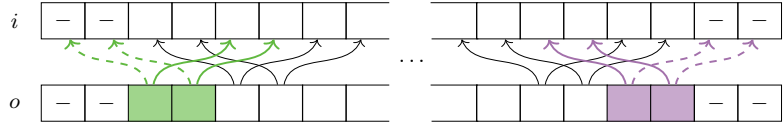
\begin{figure}[t]
  \definecolor{pre}{RGB}{33, 110, 235}
\definecolor{post}{RGB}{7, 163, 30}
\tikzset{three sided/.style={
        draw=none,
        append after command={
            [shorten <= -0.5\pgflinewidth]
            ([shift={(-1.5\pgflinewidth,-0.5\pgflinewidth)}]\tikzlastnode.north east)
        edge([shift={( 0.5\pgflinewidth,-0.5\pgflinewidth)}]\tikzlastnode.north west) 
            ([shift={( 0.5\pgflinewidth,-0.5\pgflinewidth)}]\tikzlastnode.north west)
        edge([shift={( 0.5\pgflinewidth,+0.5\pgflinewidth)}]\tikzlastnode.south west)            
            ([shift={( 0.5\pgflinewidth,+0.5\pgflinewidth)}]\tikzlastnode.south west)
        edge([shift={(-1.0\pgflinewidth,+0.5\pgflinewidth)}]\tikzlastnode.south east)
        }
    }
}
\tikzset{three sided left/.style={
        draw=none,
        append after command={
            [shorten <= -0.5\pgflinewidth]
            ([shift={(-1.5\pgflinewidth,-0.5\pgflinewidth)}]\tikzlastnode.north west)
        edge([shift={( 0.5\pgflinewidth,-0.5\pgflinewidth)}]\tikzlastnode.north east) 
            ([shift={( -0.5\pgflinewidth,-0.5\pgflinewidth)}]\tikzlastnode.north east)
        edge([shift={( -0.5\pgflinewidth,+0.5\pgflinewidth)}]\tikzlastnode.south east)            
            ([shift={( 0.5\pgflinewidth,+0.5\pgflinewidth)}]\tikzlastnode.south east)
        edge([shift={(-1.0\pgflinewidth,+0.5\pgflinewidth)}]\tikzlastnode.south west)
        }
    }
}
\begin{center}
\scalebox{0.91}{
\begin{tikzpicture}[shorten >=0pt,auto,stream/.style={draw,minimum height=15pt, minimum width=18pt},time/.style={draw=none, minimum width=18pt}]

        \node[]				(b)			at (-0.7,0)					{$i$};
        \node[]				(a)			at (-0.7,-1.2)				{$o$};

        \node[stream]		(b-1)		at (0,0)						{$-$};
        \node[stream]		(b0)			[right=-0.5 pt of b-1]		{$-$};
        \node[stream]		(b11)		[right=-0.5 pt of b0]		{};
        \node[stream]		(b12)		[right=-0.5 pt of b11]		{};
        \node[stream]		(b21)		[right=-0.5 pt of b12]		{};
        \node[stream]		(b22)		[right=-0.5 pt of b21]		{};
        \node[stream]		(b31)		[right=-0.5 pt of b22]		{};
        \node[stream,three sided]		(b32)		[right=-0.5 pt of b31]		{};
        \node[time]				(bd)			[right=-0.5 pt of b32]		{};
        \node[]				(dots)			[below=.33 of bd]		{\hspace{0.06cm}$\dots$};

        \node[stream,three sided left]		(c1)		[right=-0.5 pt of bd]		{};
        \node[stream]		(c2)		[right=-0.5 pt of c1]		{};
        \node[stream]		(c3)		[right=-0.5 pt of c2]		{};
        \node[stream]		(c4)		[right=-0.5 pt of c3]		{};
        \node[stream]		(c5)		[right=-0.5 pt of c4]		{};
        \node[stream]		(c6)		[right=-0.5 pt of c5]		{};
        \node[stream]		(c7)		[right=-0.5 pt of c6]		{$-$};
        \node[stream]		(c8)		[right=-0.5 pt of c7]		{$-$};
        
        \node[stream]		(a-1)		at (0,-1.2)					{$-$};
        \node[stream]		(a0)			[right=-0.5 pt of a-1]		{$-$};
        \node[stream]		(a11)		[fill=prefix!60,text=black,right=-0.5 pt of a0]		{};
        \node[stream]		(a12)		[fill=prefix!60,right=-0.5 pt of a11]		{};
        \node[stream]		(a21)		[right=-0.5 pt of a12]		{};
        \node[stream]		(a22)		[right=-0.5 pt of a21]		{};
        \node[stream]		(a31)		[right=-0.5 pt of a22]		{};
        \node[stream,three sided]		(a32)		[right=-0.5 pt of a31]		{};
        \node[time]				(ad)			[right=-0.5 pt of a32]		{};

        \node[stream,three sided left]		(d1)		[right=-0.5 pt of ad]		{};
        \node[stream]		(d2)		[right=-0.5 pt of d1]		{};
        \node[stream]		(d3)		[right=-0.5 pt of d2]		{};
        \node[stream]		(d4)		[right=-0.5 pt of d3]		{};
        \node[stream]		(d5)		[fill=postfix!60,right=-0.5 pt of d4]		{};
        \node[stream]		(d6)		[fill=postfix!60,right=-0.5 pt of d5]		{};
        \node[stream]		(d7)		[right=-0.5 pt of d6]		{$-$};
        \node[stream]		(d8)		[right=-0.5 pt of d7]		{$-$};
        
        \path[->]

                (a21)	edge[black,in=290,out=100]		node		{}	(b11)
                (a21)	edge[black,in=250,out=80]		node		{}	(b31)
                (a22)	edge[black,in=290,out=100]		node		{}	(b12)
                (a22)	edge[black,in=250,out=80]		node		{}	(b32)

                (d3)	edge[black,in=290,out=100]		node		{}	(c1)
                (d3)	edge[black,in=250,out=80]		node		{}	(c5)
                (d4)	edge[black,in=290,out=100]		node		{}	(c2)
                (d4)	edge[black,in=250,out=80]		node		{}	(c6)

                (a11)	edge[thick,color=prefix,dashed,in=290,out=100]		node		{}	(b-1)
                (a11)	edge[thick,color=prefix,in=250,out=80]		node		{}	(b21)
                (a12)	edge[thick,color=prefix,dashed,in=290,out=100]		node		{}	(b0)
                (a12)	edge[thick,color=prefix,in=250,out=80]		node		{}	(b22)

                (d5)	edge[thick,postfix,in=290,out=100]		node		{}	(c3)
                (d5)	edge[thick,postfix,dashed,in=250,out=80]		node		{}	(c7)
                (d6)	edge[thick,postfix,in=290,out=100]		node		{}	(c4)
                (d6)	edge[thick,postfix,dashed,in=250,out=80]		node		{}	(c8)

        ;
    \end{tikzpicture}}
\end{center}
  \caption{%
    Illustration of stream accesses in different phases of the execution.  An output stream~$o$ accesses an input stream~$i$ with offsets $-2$ and $+2$.  In the prefix (postfix) of the execution, the past (future) accesses need to be substituted by their default values.%
  }
  \label{fig:streamaccess}
\end{figure}

This behavior induces the structure of the monitor execution: it starts with a prefix where past accesses always fail, loops in the regular execution where all accesses always succeed, and ends in a postfix where future accesses always fail.

\Cref{fig:streamaccess} illustrates stream accesses in the different phases.  
It shows an output stream $o$ that accesses an input stream $i$ with an offset of $-2$ and $2$.
In the first two iterations of the monitor execution, \ie, in the prefix, the accesses to the past values will fail, requiring the monitor to use the default values instead.  
Afterwards, all accesses succeed until the input stream ends.  
In the last two evaluations, \ie, in the postfix, the future accesses fail and need to be replaced by the default values.

While the shift only concerns time, it can also be used to compute the memory requirement of a stream, \ie, the number of values of a single stream that can be relevant at the same time.
If a stream $s$ of type $T$ has a memory requirement $\memreq{s} = i$, the monitor needs to reserve $i \cdot \sizeof(T)$ bytes of memory for $s$.

\begin{definition}[Memory Requirement]
  The \emph{memory requirement} of a dependency $(s', w, s) \in E$ is determined by the shifts of the streams as well as the weight $w$ of the dependency, \ie, the offset of the stream access: $\shift{s} - \shift{s'} - w$. The memory requirement of a stream is thus the maximum requirement of any outgoing dependency:
  $\memreq{s} = \max \Set{\shift{s} - \shift{s'} - w \given (s', w, s) \in E}$.
\end{definition}

Hence, the compilation determines three key values for each specification.
\begin{definition}[Memory Consumption, Prefix- and Postfix Length]
Let $\memcon$, $\preflen$, and $\postlen$ be the \emph{memory consumption}, \emph{prefix length} and \emph{postfix length} of $\spec$, respectively, defined as follows:
\begin{flalign*}
	\quad \quad \memcon &= \sum_{s \in \spec}\Set{\memreq{s} \cdot \sizeof(T_s) } &&\\
	\preflen &= \max_{s \in \spec}\Set{\shift{s} + \memreq{s}} \\
	\postlen &= \max_{s \in \spec} \Set{\shift{s}}
\end{flalign*}
\end{definition}

Furthermore, the evaluation order $\evalorder$ of the output streams of a \lola specification induces the so-called \emph{evaluation layers}.%
\begin{definition}[Evaluation Layer]
	Let $\spec$ be a \lola specification and let $\evalorder$ be the evaluation order induced by its dependency graph. If $\layer{s} = k$ for an output stream $s$, then there is a strictly decreasing sequence of $k$ streams \wrt~$\evalorder$ starting in $s$.
\end{definition}

Intuitively, an evaluation layer consists of all streams that are incomparable according to the evaluation order. 
For the \lola specification from \Cref{fig:runningexample:spec}, for instance, the output streams \texttt{tooLow} and \texttt{tooHigh} are incomparable according to the evaluation order. Thus, they are contained in the same evaluation layer.
Evaluation Layers are also used to identify independent streams and thus to enable their concurrent evaluation as described in \cref{sec:concurrency}.

\subsection{Code Generation}

The monitor code starts with a \emph{prelude} which declares data structures and helper functions. 
It also contains the \lstinline{main} function starting with the static allocation of the working memory.
The remainder of the \lstinline{main} function is the operative monitoring code consisting of three components:  the \emph{execution prefix}, the \emph{monitor loop}, and the \emph{execution postfix}.
The general structure is illustrated in \Cref{fig:monitorstructure}, details follow in the remainder of this section.

\begin{figure}[t] 

\begin{tikzpicture}[]
	 \draw[rounded corners, fill=prelude!20, draw=prelude] (-2.1, 0) rectangle ++(6, -3.75);
	 \node[anchor=north east, inner sep=0, outer sep=0] (prelude) at (3.7, -.1) {\scriptsize\color{prelude}Prelude};
	 \draw[rounded corners, fill=loop!20, draw=loop] (-2.1, -4.3) rectangle ++(6, -1.15);
	 \node[anchor=south east, inner sep=0, outer sep=0] (loop) at (3.7, -5.35) {\scriptsize\color{loop}Monitor Loop};
	 \draw[rounded corners, fill=prefix!20, draw=prefix] (4.3, 0) rectangle ++(5.8, -3.3);
	 \node[anchor=south east, inner sep=0, outer sep=0] (pre) at (9.9, -3.2) {\scriptsize\color{prefix} Execution Prefix};
	 \draw[rounded corners, fill=postfix!20, draw=postfix] (4.3, -3.35) rectangle ++(5.8, -1.5);	 
	 \node[anchor=south east, inner sep=0, outer sep=0] (post) at (9.9, -4.75) {\scriptsize\color{postfix}Execution Postfix};
\end{tikzpicture}

\vspace{-5.75cm}
\begin{lstlisting}[,style=ColoredRust, basicstyle=\scriptsize,multicols=2, frame=none]
struct Memory { ... } $\tikzmark{preludeBegin}$
impl Memory { $\dots$ }
[[ Evaluation Functions ]]
fn get_input() -> Option<($T_{s_1}, \dots, T_{s_\ell}$)> {
  [[ Communicate with system ]]
}
fn emit(output: &($T_{s_1}, \dots, T_{s_n}$)) {
  [[ Communicate with system ]]
}
fn main() {
  let mut memory = Memory::new(); $\tikzmark{preludeEnd}$$\tikzmark{longest-pre}$
  let early_exit = prefix(&mem);
  if !early_exit {
    while let Some(input) = get_input() { $\tikzmark{longest}$$\tikzmark{loopBegin}$
	  mem.add_input(&input1);
      [[ Evaluation Logic ]]
    } $\tikzmark{loopEnd}$
  }
  postfix(&mem);
}
fn prefix(mem: &mut Memory) -> bool {
 if let Some(input) = get_input() {$\tikzmark{prefixBegin}$
   mem.add_input(&input);
	  [[ Evaluation Logic ]]
	} else {
     return true // Jump to Postfix.
	}
	[[ Repeat $\color[rgb]{0.6, 0.4, 0.08}\preflen$ times. ]] $\tikzmark{prefixEnd}$
	false // Continue with Monitor Loop.
}

fn postfix(mem &Memory) {
 [[ Evaluation Logic ]] $\tikzmark{postfixBegin}$
 [[ Repeat $\color[rgb]{0.6, 0.4, 0.08}\postlen$ times. ]] $\tikzmark{postfixEnd}$
}
$~$
$~$
$~$
$~$
$~$
$~$
\end{lstlisting}
\tikzset{
  line width=0.2pt
}
\captionof{lstlisting}{Structure of the generated \rust code.}\label{fig:monitorstructure}
\end{figure}

\paragraph*{Prelude.}
The prelude declares several functions required throughout the monitor execution and declares as well as allocates the working memory.
The functions consist of two I/O functions and evaluation functions. 

\newcommand*\texpadsuxone{T_{s_1}}
\newcommand*\texpadsuxtwo{T_{s_\ell}}
The \lstinline{get_input() -> Option<($\texpadsuxone,\dots,\texpadsuxtwo$)>} function, where $T_{s_1},\dots,T_{s_\ell}$ are the types of all input streams, models the receipt of input data.
It produces either \lstinline{None} if the execution of the system under scrutiny terminated, or \lstinline{Some(v)}, where $v$ is an $\ell$-tuple containing the latest input values.
\newcommand*\texpadsuxthree{T_{s_\ellplusone}}
\newcommand*\texpadsuxfour{T_{s_k}}
Conversely, the function \lstinline{emit(&($\texpadsuxthree,\dots,\texpadsuxfour$))} conveys a $(k-\ell)$-tuple of output values to the system.

For each stream, there are evaluation functions in several variants depending on whether they will be called in the prefix, the loop, or the postfix.
The implementations differ only in the logic accessing other streams.  
The \lola semantics dictates that the evaluation needs to check whether the accessed value exists and to substitute it with the respective default value if needed.
However, an analysis of the dependency graph reveals statically which accesses will fail.
Thus, providing several implementations makes the need for such a check during runtime redundant.

The working memory is a struct aptly named \lstinline{Memory}.
It consists of a static array for each stream in the specification and reads as follows:
\begin{lstlisting}[style=ColoredRust]
struct Memory { $s_1$: [$T_{s_1}$, $\memreq{s_1}$], $\dots$ , $s_k$: [$T_{s_n}$, $\memreq{s_n}$] }  
\end{lstlisting}
Here, $s_1, \dots, s_k$ are all input and output streams with types $T_1,\dots,T_k$.
The monitor allocates \lstinline{Memory} once in its main function, keeps it on the stack, and grants read access to functions evaluating stream expressions.

\paragraph*{Execution Prefix.}
The prefix consists of $\preflen$ conditional blocks, each processing an input event of the system under scrutiny. 
If the system terminates before the prefix concludes, the function returns true, indicating an early termination, which prompts the \lstinline{main} function to initiate the postfix.
Otherwise, the input is added to the working memory and, evaluation layer by evaluation layer, each output stream is evaluated in a dedicated function as can be seen in the following code snippet.  
For this, assume that the specification has $\lambda^\ast$ evaluation layers, \ie, 
$ 
  \lambda^\ast = \max \Set{x \given \exists s_1, \dots, s_x\colon s_1 \evalorder \dots \evalorder s_x} 
$ 
Moreover, $\lambda_i = \card{\Set{s \given \layer{s} = i}}$ denotes the number of streams within evaluation layer $i \leq \lambda^\ast$.
Lastly, let $s_{i,j} \extevalorder s_{i, j+1}$ with $\layer{s_{i,j}} = \layer{s_{i,j+1}} = i$.

\begin{lstlisting}[style=ColoredRust]
  let val_$s_{1,1}$ = eval_pre_1_$s_{1,1}$(&Memory);
  ...
  let val_$s_{1,\lambda_1}$ = eval_pre_1_$s_{1,\lambda_1}$(&Memory); 
  memory.write_layer_1(val_$s_{1,1}$, ..., val_$s_{1, \lambda_1}$)
  ...
  let val_$s_{\lambda^\ast,1}$ = eval_pre_$s_{\lambda^\ast,1}$(&Memory);
  ...
  let val_$s_{\lambda^\ast,\lambda_{\lambda^\ast}}$ = eval_pre_$s_{\lambda^\ast,\lambda_{\lambda^\ast}}$(&Memory);
  Memory.write_layer_$\lambda^\ast$(val_$s_{\lambda^\ast,1}$, ..., val_$s_{\lambda^\ast, \lambda_{\lambda^\ast}}$);$\tikzmark{longest}$
  if val_$s_{t_1}$ == true { emit($m_{t_1}$) }
\end{lstlisting}

Note that, as indicated in the prelude, each conditional block calls a different set of evaluation functions.
This allows for a fine-grained treatment of stream accesses, improving the overall performance at the cost of greater code size.
Also, the call passes a single argument to the evaluation function: an immutable reference for \lstinline{Memory}.
As a result, the \rust type system guarantees that the evaluation does not mutate its state.
The function returns a value that is committed to \lstinline{Memory} after fully evaluating the current layer.
The bodies of these functions are straight-forward translations of stream expressions: each arithmetic and logical expression has a counterpart in \rust.  
Stream lookups access the only argument passed to the function, \ie, a read-only reference to the working memory.

The \lstinline{write_layer_$i$} functions commit computed stream values to \lstinline{Memory}. 
After $\memreq{s}$ iterations, the memory evicts the oldest data point for stream $s$, thus constituting a ring buffer.

\paragraph*{Monitor Loop}
The main difference between the monitor loop and the prefix is, as the name indicates, that the former consists of a loop.
The loop terminates as soon as the system ceases to produce new inputs.
At this point, the monitor transitions to the execution postfix.

Within the loop, the monitor proceeds just as in the prefix except that the evaluation functions are agnostic to the current iteration number.
In the evaluation, all stream accesses are guaranteed to succeed rendering the evaluation free of conditionals except when the stream expression itself contains one.

\paragraph*{Execution Postfix}
The structure of the execution postfix closely resembles the prefix except for two differences: The postfix does not check for the presence of new input values and calls a different set of evaluation functions, specifically tailored for the postfix iteration.

\paragraph*{Code Characteristics}
The generated code exhibits two advantageous characteristics.  
First, the trade-off between an increase in code size by quasi-duplicating the evaluation functions leads to an excellent performance in terms of running time. 
The functions require few arguments, avoid conditional statements as much as possible, and utilize memory locality. 
This is further emphasized by the lack of dynamic memory allocation and utilization of native datatypes.
Second, the clear code structure, especially \wrt memory accesses, drastically simplifies reasoning about the correctness of the code.

\section{Verification}\label{sec:verificationannotations}

Our goal is to prove that the verdicts produced by the monitor correspond to the formal semantics.
The main challenge is that the the evaluation model of the \lola semantics refers to unbounded data sequences, disregarding any memory concerns.
The implementation, however, manages the monitoring process with only a finite amount of memory.
As a result, the \lola semantics may refer to data values long after they have been discarded in the implementation.
Hence, the relation between the memory content and the evaluation model, and thus the correctness of the computation, is no longer apparent.

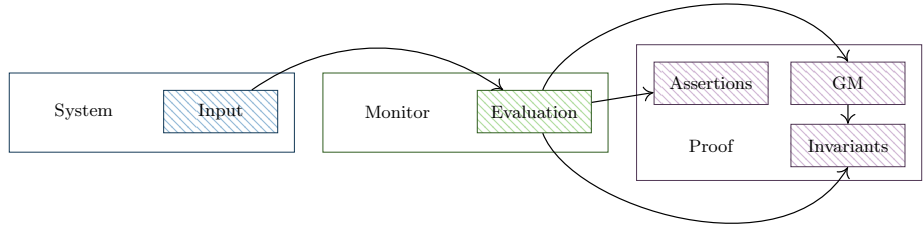
\begin{figure}[t]
\centering
  \scalebox{.75}{
\usetikzlibrary{calc}
  \begin{tikzpicture}[yscale=-1]
    \tikzstyle{component}=[draw, rectangle, minimum height=.75cm, minimum width=2cm, anchor=north west, pattern=north west lines]
    \tikzstyle{module}=[minimum height=.75cm, minimum width=2cm, anchor=north west]
    \tikzstyle{flow}=[->, semithick]
    \tikzstyle{system}=[pattern color=loop!50!white, draw=loop!50!black]
    \tikzstyle{monitor}=[pattern color=prefix!50!white, draw=prefix!50!black]
    \tikzstyle{proof}=[pattern color=postfix!50!white, draw=postfix!50!black]
    
    \node[component, system] (input) at (2.7, .3) {Input};
    \node[module, system, draw=none] (system) at (.3, .3) {System}; 
    \draw[system] (0, 0) rectangle ++(5.0, 1.4); 
    
    \node[component, monitor] (eval) at (8.2, .3) {Evaluation};
    \node[module, monitor, draw=none] (monitor) at (5.8, .3) {Monitor};
    \draw[monitor] (5.5, 0) rectangle ++(5.0, 1.4);

    \node[component, proof] (gm) at (13.7, -0.2) {GM}; 
    \node[component, proof] (assert) at (11.3, -0.2) {Assertions};
    \node[component, proof] (inv) at (13.7, 0.9) {Invariants};
    \node[module, proof, draw=none] (proof) at (11.3, 0.9) {Proof};
    \draw[proof] (11, -0.5) rectangle ++(5.0, 2.4);

    \draw[flow,-{>[scale=2.5,length=2,width=3]}] (input) to[bend right=35] (eval);
    \draw[flow,-{>[scale=2.5,length=2,width=3]}] (eval) to[out=285,in=240] (gm.north);
    \draw[flow,-{>[scale=2.5,length=2,width=3]}] (eval) to[out=75, in=120] (inv.south);
    \draw[flow,-{>[scale=2.5,length=2,width=3]}] (gm) edge (inv);
    \draw[flow,-{>[scale=2.5,length=2,width=3]}] (eval) edge (assert);
    
  \end{tikzpicture}
}
  \vspace{-0.6cm}
  \caption{Information flow between the monitor and the ghost memory.}
  \label{fig:verification:flow}
\end{figure}

We solve this problem with the classic proof technique of introducing so-called \emph{ghost memory}.
The compilation introduces another data structure named \lstinline{Ghost Memory} (GM) which is a wrapper for \rust vectors, \ie, dynamically growing sequences of data.
Whenever the monitor receives or computes any data, it commits it to the GM.
The GM's size thus obviously exceeds any bound, voiding the memory guarantees.
However, the ghost memory's sole purpose is to aid the verification and not the monitor; information flows from the program into the GM and the proof, but remains strictly separated from the monitor execution.
This allows for removing the GM after successfully verifying the correctness of the monitor without altering its behavior.
\Cref{fig:verification:flow} illustrates the flow of information between the monitor and GM.
Clearly, the monitor remains unaffected when removing any proof artifacts.

The correctness proof has two major obligations: proving compliance between values in the GM and the working memory, and proving the correctness of the trigger evaluations \wrt the ghost memory.
These obligations are encoded as verification annotations, such that the \viper framework verifies them automatically.
The compilation generates additional annotations to guide the verification process.
\viper annotations fall into the following categories:
\begin{description}
\item[Function Contracts] 
  Annotations in front of a function \lstinline{f} consist of preconditions and guarantees.
  \viper imposes constraints on the function caller and the function body itself.
  Each call to \lstinline{f} is replaced by an assertion of the preconditions of \lstinline{f}, prompting \viper to prove their validity, and an assumption of the guarantees. 
  In a separate step, \viper assumes the preconditions and verifies that the guarantees hold after executing the function body.
  Note that the \rust type system already ensures that references passed to the function are accessible and cannot be modified or freed unless they are explicitly declared mutable.
\item[Loop Invariants]
  \viper analyzes while-loops similarly to functions in three steps.
  First, the code leading to the loop needs to satisfy the invariants.
  Second, \viper assumes both the loop invariant and the loop condition to hold and verifies that the invariant again holds after the execution of the body.
  Lastly, \viper assumes the invariant and the negation of the loop condition to hold for the code after the loop.
\item[Inline Assertions]  
  Both loop invariants and function contracts impose implicit assertions on the code.
  \viper allows for supplementing them with explicit inline assertions using the \rust \lstinline{assert!} macro. 
  Usually, the macro checks an expression during runtime.
  \viper, however, eliminates the need for this dynamic check as it verifies the correctness statically and transforms it into an assumption for the remainder of the verification.
  Thus, the assertions serve a similar function as the ghost memory: they are a proof construct and do not influence the monitor per se (\cf \Cref{fig:verification:flow}).
\end{description}

\paragraph*{Annotation Generation.}
The compilation inserts annotations at several key locations.
First, as an example for function annotations, consider a function that retrieves a value of the stream~$s$ from the working memory.
The function takes the relative index of the retrieved value as single argument, \ie, an index of~$1$ accesses the second to newest value.
The annotation requires that the index must not exceed the memory reserved for~$s$.
Syntactically, this results in the following annotation in front of the function head: \lstinline{#[requires="index < $\lstmathanno\memreqs$"]}.
Moreover, the function needs to guarantee that the return value corresponds to the respective value stored in \lstinline{Memory}. This is expressed by the annotation \lstinline{#[ensures="index == $\lstmathanno i$ ==> result == self.$\lstmathanno s$[$\lstmathanno i$]"]} for each $i \leq \memreq{s}$.
The remaining function annotations follow a similar pattern, \ie, they require valid arguments, and ensure correct outputs as well as the absence of undesired changes.
Note that the ghost memory is essentially a wrapper for \rust vectors as they represent a growing list of values.
Thus, functions concerning the ghost memory carry the standard annotation ensuring correctness of the vector as presented in the \viper examples.\footnote{See \eg the verified solution for the Knapsack Problem: \url{https://github.com/viperproject/prusti-dev/blob/master/prusti/tests/verify/pass/rosetta/Knapsack_Problem.rs}.}

Second, the loop has several entry checks that are expressed as inline assertions.
These ensure that the iteration count is $\preflen$ and that the length of the ghost memory for a stream $s$ is $\preflen - \shift{s}$.
This is necessary because the loop invariant asserts equivalence between an excerpt of the ghost memory and the working memory.
While the existence of all accessed values in the working memory is guaranteed due to the static allocation, the GM grows dynamically.
Hence, the compilation adds the entry checks.

In terms of memory equivalence, it remains to be shown that all values in the working memory correspond to the respective entry in the ghost memory.
Formally, let $m$ be the working memory and let $g$ be the ghost memory where index 0 marks the latest value.  
Furthermore, let $\eta$ be the current iteration count.
Then, the invariant checks:
\begin{equation}\label{eq:maininv}
  \forall s\colon\forall i\colon (0 \leq i < \memreq{s} \implies m_s[i] = g_s[i]).
\end{equation}
At loop entry, $\memreq{s} = \preflen - \shift{s} = \eta - \shift{s}$ is the number of iterations in which a value for $s$ was computed.
In each further iteration of the loop, the invariant checks that the former $\memreq{s}-1$ entries remained the same and that the new values in the ghost memory $g$ and the working memory $m$ are equal.
The first of these checks is not strictly necessary for the proof because it immediately follows from the function contracts of the helper functions.
However, after completing one loop iteration, \viper deletes prior knowledge about all variables that were mutated in the loop. 
Further reasoning about these variables is thus solely based on the loop invariants.

To express \Cref{eq:maininv} in \viper, the compilation needs to statically resolve the universal quantification over the streams.
Thus, for each stream~$s$, the compilation generates the annotation
\lstinline{#[invariant="forall i: usize :: (0 <= i && i < $\lstmathanno\memreqs$) ==> mem.get_$\lstmathanno s$(i) == gm.get_$\lstmathanno s$(iter - 1)"]},
where \lstinline{iter} is a variable denoting the current iteration, \lstinline{mem} is the working memory, and \lstinline{gm} is the ghost memory.
\viper is able to handle the remaining universal quantification over $i$.
However, the compilation reduces the verification effort further by unrolling it. This is possible since the memory requirement $\memreq{s}$ of a stream $s$ is determined statically.

Lastly, the compilation introduces inline assertions after the evaluation of stream expressions, \ie, in the prefix, postfix, and loop body.
These annotations show that computed values are correct when assuming that the values retrieved from the working memory are correct as well.
This argument is well-founded because the compilation substitutes failing stream accesses by their respective default values.
Thus, any value retrieved from \lstinline{Memory} was computed in an earlier iteration or layer and therefore proven correct by \viper.

It only remains to be shown that the stream expression is properly evaluated. 
Expressions consist of arithmetic or logical functions, constants, and stream accesses.
The former two can be trivially represented in \viper.
Since the memory is assumed to be correct and failing accesses are substituted by constants when possible, accesses also translate naturally into \viper.

\paragraph*{Conclusion.}
The validity of the assertions after the evaluation logic shows that newly computed values are correct if the values in the working memory $m$ and the ghost memory $g$ coincide.
This fact is guaranteed by the loop invariant.
Furthermore, the inductive argument of the loop invariants allows us to conclude that, if $m$ were to never discard values, $m_s[i] = g_s[i]$ for all streams $s$ and $i \leq \eta$.
Thus, $m$ is a real subsequence of $g$, which is a perfect reflection of the evaluation model.
As a result, any trigger violation detected by the monitor realization corresponds to a violation in the evaluation model for the same sequence of input values; The realization is verifiably correct.

\section{Concurrent Evaluation}\label{sec:concurrency}

Evaluating independent streams concurrently can significantly improve the performance of the monitor. In the following, we devise an analysis of \lola specifications that enables safe parallelization.
We observe two characteristics of \lola:
the computation of a stream expression can only \emph{read} the memory of other streams, and inter-stream dependencies are determined statically. 
The evaluation layers are a manifestation of the second observation.
They group streams which are incomparable according to the evaluation order.
Combined with the first observation, we can conclude that all streams within one layer may be computed in parallel.
Thus, the compilation spawns a new thread for each stream within the layer with read access to the global memory. 
We add annotations to the code that enable \viper to verify that the parallel execution remains correct.

The compilation capitalizes on \rust's concurrency capabilities by evaluating different output streams in parallel.
A major advantage of \rust is that its ownership model enforces a strict separation of mutable and immutable data.
Any data point has exactly one owner who can transfer ownership for good or let other functions borrow the data.
Borrowing data is again either mutable or immutable.
If a function mutably borrows data, no other function, including the owner, can read or write this data.
Similarly, if a function immutably borrows data, other functions and the owner can only read it.
A consequence of this fine-grained access management with static enforcement is that enabling concurrency becomes rather easy when compared to languages like C.

Enabling the concurrent evaluation requires slight changes in the code generation.
First, evaluation functions are annotated with \lstinline{#[pure]}.
This indicates that a function mutates nothing but its local stack portion.
For the evaluation logic, the compiler still proceeds layer by layer, opening a \emph{scope} for each of them.
In the scope, it generates code following the total evaluation order $\extevalorder$.
However, rather than calling the respective evaluation functions directly, the parallelized version spawns a thread for each stream and starts the evaluation inside it.
Assume $s_1, \dots, s_n$ constitute a single layer of a specification.
The evaluation then looks as follows:
\begin{lstlisting}[style=ColoredRust]
let (v_1, ..., v_n) = crossbeam::scope(|scope| {
  let handle_s1 = scope.spawn(move |_| {
      eval_s1(&memory)
  });
  ...
  let handle_sn = scope.spawn(move |_| {
      eval_sn(&memory)
  });
  (handle_s1.join().unwrap(), ..., handle_sn.join().unwrap())
}).unwrap()
\end{lstlisting}

Note that the code snippet uses the \rust crate crossbeam, a standard concurrency library. 
A similar result can be achieved without external code by moving the global memory to the heap and using the standard \rust thread logic.\footnote{%
On a technical note: \rust's type system requires the programmer to guarantee that the global memory will not be dropped until all threads terminate.  
Thus, the memory needs to be wrapped into an \emph{Atomically Reference Counted (Arc)} pointer.  
This has two disadvantages:  all accesses to memory require generally slower heap access and the evaluation suffers from the overhead accompanying atomic reference counting.}

The correctness of this approach is an immediate consequence of the correctness of the evaluation order and memory locality of streams.
In particular, the independence of streams within the same evaluation layer and the pureness of the functions are crucial.
The latter ensures that the function does not mutate anything outside of its local stack.
The former ensures that using pure evaluation functions within the same layer is indeed possible.  
Thus, the order of execution cannot change the outcome of the function, enabling the concurrent evaluation.

Note that spawning a thread for each stream evaluation is a double-edged sword. 
While it can drastically reduce the monitor's latency, each spawn induces a constant overhead.
Thus, reducing the number of spawns while increasing the parallel computation time maximizes the gain.
Consequently, the monitor benefits stronger from the parallel evaluation when its dependency graph is wide, enabling several cores to compute in parallel.
Similarly, specifications with large stream expressions benefit from the multi-threading because the share of parallel computations increases. 
This lowers the relative impact of the constant thread-spawning overhead.

\section{Experimental Evaluation}\label{sec:casestudy}

The implementation of the compiler is based on the \rtlola\footnote{\url{http://www.rtlola.org/}} framework written in \rust.
The code verification uses the \rust-frontend of the \viper framework called \prusti~\cite{prusti}.
\prusti translates a \rust program into the \viper intermediate verification language, followed by a translation into an \textsc{smt} model, which is checked by the Z3~\cite{z3} \textsc{smt} solver.
Thus, our toolchain enables completely automatic proof checking.

The experiments were conducted on a machine with a $3.1\giga\hertz$ Dual-Core Intel i5 processor with $16\giga\byte$ of \textsc{ram}.
The artifacts for the evaluation are available on github.\footnote{\url{https://github.com/reactive-systems/Lola2RustArtifact}}
In all experiments, the compilation itself has a negligible running time of under ten milliseconds and memory consumption of less than 4\mega\byte, mainly due to the \rtlola frontend.
As expected, the verification of the annotated rust code using \prusti and the \viper toolkit takes significant time and memory.
While the translation into the \textsc{smt} model is deterministic and can be parallelized, the verification with Z3 exhibits generally high and unpredictable running time.

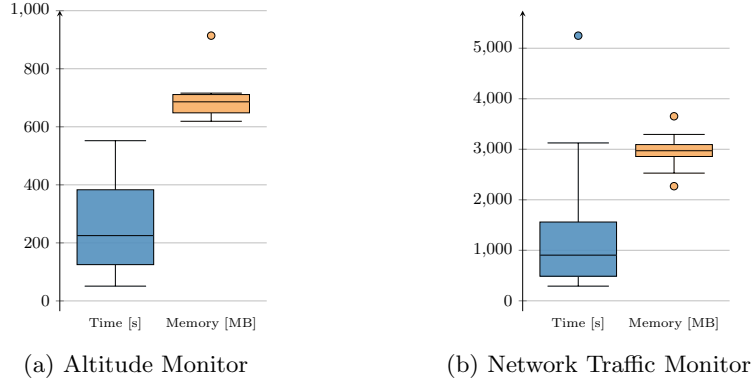
\begin{figure}[t]
	\centering
	\pgfplotsset{compat = 1.15, cycle list/Set1-4} 
\begin{subfigure}[t]{.49\textwidth}
  \centering
  \scalebox{0.7}{
  \begin{tikzpicture}
    \begin{axis}[
      boxplot/draw direction = y,
      x=1.8cm,
      x axis line style = {opacity=0},
      axis x line* = bottom,
      axis y line = left,
      enlarge y limits,
      enlarge x limits=0.1,
      ymajorgrids,
      xtick = {1, 2},
      xticklabel style = {align=center, font=\scriptsize,rotate=0},
      xticklabels = {Time [s], Memory [MB]},
      xtick style = {draw=none}, 
    ]
    \addplot+[draw=black, 
	fill = loop!80,
    fill opacity=0.9,
    boxplot prepared={
    lower whisker=51,
     lower quartile=125,
     median=225,
     upper quartile=383,
     upper whisker=552
    }
   ] coordinates {}; 
  
   \addplot+[draw=black, 
	fill = prelude!80,
    fill opacity=0.9,
    boxplot prepared={
     lower whisker=619,
     lower quartile=648,
     median=686,
     upper quartile=711 ,
     upper whisker=716
     }
   ] coordinates {(0,914)}; 
  
  \end{axis}
  \end{tikzpicture}
  } 
  \subcaption{Altitude Monitor}
\label{fig:eval:motivating}
\end{subfigure}
\begin{subfigure}[t]{.49\textwidth}
  \centering
  \scalebox{0.7}{
\begin{tikzpicture}
  \begin{axis}[
    boxplot/draw direction = y,
    x=1.8cm,
		x axis line style = {opacity=0},
		axis x line* = bottom,
    axis y line = left,
    enlarge y limits,
    enlarge x limits=0.1,
		ymajorgrids,
		xtick = {1, 2},
		xticklabel style = {align=center, font=\scriptsize,rotate=0},
		xticklabels = {Time [s], Memory [MB]},
		xtick style = {draw=none}, 
  ]
  \addplot+[draw=black, 
  fill = loop!80,
  fill opacity=0.9,
  boxplot prepared={
    lower whisker=291,
    lower quartile=487,
    median=904,
    upper quartile=1560,
    upper whisker=3125
  }
 ] coordinates {(0,5249)}; 

 \addplot+[draw=black,  
  fill = prelude!80,
  fill opacity=0.9,
  boxplot prepared={
   lower whisker=2528,
   lower quartile=2857,
   median=2971,
   upper quartile=3093,
   upper whisker=3294
   }
 ] coordinates {(0,2269)  (0,3653)}; 

\end{axis}
\end{tikzpicture}
  } 
\subcaption{Network Traffic Monitor}
\label{fig:eval:network}
\end{subfigure}%
  
	\vspace{0.2cm}
	\caption{Results of $20$ runs in terms of running time (blue, in seconds) and memory consumption (orange, in MB) for the verification of the annotated \rust code of the specification, where the altitude of a drone is monitored (cf. \Cref{fig:runningexample:spec}), and the network traffic monitor specification.}
\end{figure}

We discuss the results of compiling and verifying three \lola specification of varying size.
The process works flawlessly on two of them while the third one occasionally runs into timeouts and inconclusive verification results.

First, we consider the specification from \Cref{fig:runningexample:spec}, where the altitude of a drone is monitored. The results in terms of both running time and memory consumption for 20 runs are depicted in \Cref{fig:eval:motivating}.
Note that the y-axis displays both the running time in seconds (left plot) and the memory consumption in megabytes (right plot).
The plot shows that the running time never exceeds $600\second$ with a median of $225\second$.
The memory consumption is significantly more stable ranging between $648$ and $711\mega\byte$ with one outlier ($914\mega\byte$).

While the first specification was short and illustrative, the second one is more practically relevant.
The specification monitors the network traffic of a server based on the source and destination IP of requests, \textsc{tcp} flags, and the length of the payload~\cite{fpgalola}.
The specification counts the number of incoming connections and computes the workload, \ie, the number of bytes received over push requests.
If any of these numbers exceeds a threshold, the specification raises an alarm.
Moreover, it keeps track of the number of open connections.
A trigger indicates when the the server attempts to close a connection even though none is open.
The full specification can be found in \Cref{fig:networkspec}.
\Cref{fig:eval:network} depicts the results both in terms of running time and memory consumption for $20$ runs.
Again, the y-axis represents both running time in seconds and memory consumption in megabytes.
The increase in resource consumption clearly reflects the increase in complexity and size of the input specification.
While the longest run took nearly $90\minute$, most of the runs took less than $25\minute$ with a median of roughly $15\minute$.
Like before, the memory consumption is relatively stable ranging around $3\giga\byte$.

\begin{lstlisting}[
  style=LolaDefault,
  float,
  floatplacement=t,
  caption=\lola specification for monitoring network traffic,
  label=fig:networkspec,
  basicstyle=\ttfamily\scriptsize,
]
input src, dst, length: Int32
input fin: Bool, push: Bool, syn: Bool
constant server: Int32 := ...

output count : Int32 := if count[-1,0] > 201 then 0 else count[-1,0] + 1
output receiver : Int32 := if dst=server then receiver[-2,0] + 2 else 
    if count > 200 then 0 else receiver[-1,0]
trigger receiver > 50 "Many incoming connections."

output received : Int32 := if dst=server $\land$ push then 0 else length
output workload : Int32 := if count > 200 then workload[-1,0] + 1 else 0
trigger workload > 25 "Workload too high."
output opened : Int32 := opened[-1,0] + int(dst=server $\land$ syn)
output closed : Int32 := closed[-1,0] + int(dst=server $\land$ fin)
trigger opened - closed < 0 "Closed more connections than have been opened."
\end{lstlisting}

Lastly, we considered a \lola specifications that shows the limitations of our approach.
It detects different flight phases of a drone and raises an alarm if actual velocity and a reference velocity provided by the flight controller deviate strongly.
The specification is based on a \lola specification for flight phase detection shown in \Cref{fig:flightphasespec}.

\begin{lstlisting}[
style=LolaDefault,
caption=\lola specification for flight phase detection,
label=fig:flightphasespec,
float,
floatplacement=t,
basicstyle=\ttfamily\scriptsize,
]
input time_s, time_micros, velo_x, velo_y, velo_r_x, velo_r_y: Int32

output time := time_s + time_micros / 1000000
output count := count[-1,0] + 1
output frequency := 1 / (time - time[-1,0])
output freq_sum := frequency + freq_sum[-1,0]
output freq_avg := freq_sum / count
output velo : Int32 := vel_x*vel_x + vel_y*vel_y
output velo_max : Int32 := if res_max[-1,false] then velo 
    else max(velo_max[-1,0], velo)
output velo_min : Int32 := if res_max[-1,false] then velo 
    else min(velo_min[-1,0], velo)
output res_max: Bool := (velo_max - velo_min) > 1
output unchanged: Int32 := if res_max[-1,false] then 0 else unchanged[-1,0] + 1
output velo_dev : Int32 := velo_r_x - velo_x + velo_r_y - velo_y
output worst_dev: Int32 := if unchanged > 15 then velo_dev else max(velo_dev, worst_dev[-1,-10])

trigger freq_avg < 10 "Low input frequency."
trigger velo_dev > 10 "Deviation between velocities too high."
trigger worst_dev > 20 "Worst velocity deviation too high."
\end{lstlisting}

%
%
%
%

After a successful compilation, the verification was able to reveal potential arithmetic errors in the original specification~\cite{uav2}.
The errors arose from division in which the denominator was an input stream access.
The resulting value is not necessarily non-zero, so \viper reported that the respective annotation cannot be verified.
Hence, our approach is able to detect flaws in specifications stemming from implicit assumptions on the system.
These assumptions may not hold during runtime, causing the monitor to fail.

Thus, we modified the flight phase detection specification to work without division.
Yet, only four of our runs terminated successfully. The running time varies between $6$ and $16\minute$ and the memory consumption between $1.38\giga\byte$ and $1.66\giga\byte$.
The successful runs show that our approach is able to verify monitor realizations of large and arithmetically challenging \lola specifications.
However, two runs did not terminate within three hours.
The reason lies within the underlying \textsc{smt} solver: an unfavorable path choice in the solving procedure can result in extended running times.
Additionally, for four runs, the verification reported that some assertions might not hold or crashed internally.
While restarting the verification procedure can lead to finding a successful run, the incident shows the reliance of our approach on external tools.
Hence, the applicability increases with advances in research on automated proof checking of annotated code.
This constitutes another reason for the continued development of valuable tools like \prusti and the \viper framework.

\subsection{Performance of Generated Monitors}
As expected, the compiled monitors exhibit superior running time when compared against the \rtlola~\cite{rtlolacavtoolpaper} interpreter.
The comparison is based on randomly generated input data for the Altimeter\footnote{The specification was adapted to be compliant with \rtlola: rather than accessing the input with a future offset, the specification used a negative offset of -2.} and Network Traffic Monitor.
For the first specification, the interpreter required $438\nano\second$ per event on average out of 10 runs, whereas the compiled version took $6.2\nano\second$.  
The second, more involved specification shows similar results: $1,535\micro\second$ for the interpreter and $63.4\nano\second$ for the compiled version.
\section{Related Work}\label{sec:rw}

The development of a verifying compiler was identified by Tony Hoare as a grand challenge for computing research~\cite{10.1007/978-3-540-45213-3_4}.
Milestone results have been the concept of proof-carrying code (\textsc{pcc})~\cite{proofcarryingcode} and the technique of checking the result of each compilation instead of verifying the compiler's source code~\cite{certifyingcompiler}.
\textsc{pcc} architectures~\cite{pccjava} and certifying compilers~\cite{ccjava} exist for general purpose languages like \java.
A variation of the \textsc{pcc}, abstraction-carrying code~\cite{acc, cacc} was developed for constraint logic programs, where a fixpoint of an abstract interpretation serves as certificate for invariants. 
This enables automatic proof generation.

In this paper, we present a verifying compiler for the stream-based monitoring language \lola. Compared to general programming languages, the compilation of monitoring languages is still a young research topic. 
Some work has focused on compiling specifications immediately into executable code.
Rmor~\cite{rmor}, for instance, generates constant memory C code.

Similarly, a \copilot~\cite{copilot} specification can be compiled into a constant memory and constant time C realization.
The \copilot toolchain~\cite{copilotembedded} enables the verification of the monitor using the \textsc{cbmc} model checker~\cite{cbmc}.
As opposed to our approach, their verification is limited to the absence of various arithmetic errors, lacking functional correctness.
While \textsc{cbmc} can verify arbitrary inline assertions, \copilot does not generate them.
Note that, in contrast to \lola, \copilot can express real-time properties.

\rtlola\cite{rtlolaarxiv,maxmaster}, on the other hand, is a real-time, asynchronous extension of \lola, for which a compilation into the hardware description language~\vhdl exists~\cite{fpgalola}.
The \vhdl code contains traceability annotations~\cite{janmaster} and can then be realized on an \fpga.
Similarly, Pellizzoni~\etal~\cite{pellizzioni} and Schumann~\etal\cite{rtutjournal,rtutrv} realize their runtime monitors on \fpga{}s, yet without verification or traceability.

Rather than using a dedicated specification language, there are several logics for which verified compilers exist.
Differential dynamic logic~\cite{ddl}, for example, was specifically designed to capture the complex hybrid dynamics of cyber-physical systems.
The ModelPlex~\cite{modelplex} framework translates such a specification into several verified components monitoring both the environment \wrt the assumed model and the controller decisions.
Lastly, there is work on verifying monitors for metric first-order temporal~\cite{metriccompiler} and dynamic logic\cite{metriccompiler2}. 

%

\section{Conclusion}\label{sec:conclusion}

We have presented a compilation of \lola specifications into \rust code. 
Using \rust as the compilation target has the advantage that the executables are highly performant and can be used directly on many embedded platforms.
The generated code contains annotations that enable the verification of the code using the \viper framework.
With the guiding assertions in the code, as well as function contracts and loop invariants, \viper can verify monitors even for large specifications.

Our results are promising and encourage further research in this direction, such as compiling more expressive dialects of \lola such as \rtlola~\cite{rtlolaarxiv,maxmaster}.
\rtlola extends \lola with real-time aspects and can handle asynchronous inputs.
The added functionality is highly relevant in the design of monitors for cyber-physical systems~\cite{rtlolacavindustrial,rtlolacavtoolpaper}. 
While generating verifiable \rtlola monitors in \rust will require additional effort, such an extension would further improve the practical applicability of our approach.

\bibliographystyle{splncs04}
\bibliography{bibliography}

\end{document}